\documentclass[sigconf,natbib=false]{acmart}
\AtBeginDocument{%
  }
    
\setcopyright{acmlicensed}

\usepackage{xspace}
\usepackage{tikz}
\usetikzlibrary{arrows.meta,positioning,shapes.geometric,calc,fit,backgrounds}
\usepackage{graphicx}
\usepackage{enumitem}
\usepackage{booktabs}

\usepackage{array}
\usepackage{tabularx}
\usepackage{pifont}
\usepackage{colortbl}
\usepackage{pgfplots}
\pgfplotsset{compat=1.18}
\usepgfplotslibrary{groupplots}
\RequirePackage[datamodel=acmdatamodel,style=acmnumeric]{biblatex}
\newcommand{\schemename}{\textsc{SensorGuard-RV}\xspace}
\setcopyright{acmlicensed}
\copyrightyear{2027}
\acmYear{2027}
\acmDOI{XXXXXXX.XXXXXXX}
\acmConference[SAC'27]{The 42nd ACM/SIGAPP Symposium on Applied Computing}{April 5--9, 2027}{Gwangju, South Korea}
\acmISBN{979-X-XXXX-XXXX-X/27/04}

\author{Hamid Noori}
\affiliation{%
  \department{Department of Computer Science}
  \institution{Durham University}
  \city{Durham}
  \country{United Kingdom}
}
\email{hamid.noori@durham.ac.uk}

\author{Carlton Shepherd}
\affiliation{%
  \department{Department of Computer Science}
  \institution{Durham University}
  \city{Durham}
  \country{United Kingdom}
}
\email{carlton.g.shepherd@durham.ac.uk}

\usepackage{hyperref}

\begin{document}

\title[SensorGuard-RV: Sensor-to-Register Data Protection]{SensorGuard-RV: Sensor-to-Register Data Protection for Resource-Constrained RISC-V Devices}

\renewcommand{\shorttitle}{SensorGuard-RV: Sensor-to-Register Data Protection}

\begin{abstract}
Sensor readings in embedded systems pass through a hardware-firmware stack where an attacker can intercept, overwrite, or fabricate them. At worst, this may cause downstream systems, e.g., flight and industrial controllers, to behave unsafely. Existing defences, e.g., TEEs and trusted I/O paths, depend heavily on trusted software and correctly configured access control, which remain fragile in practice. In response, we present SensorGuard-RV, a hardware-first approach that delivers sensor samples into processor registers without trusting intermediate software, buses, or memory. After configuration, a write-once ownership switch hands the sensor to a fixed-function controller that performs hardware-level authenticated encryption of sensor samples under a key that is inaccessible to software. We develop a new RISC-V instruction that directs the CPU to decrypt and write the sample to a register only after verifying its tag and rejecting replays, including across resets. We implemented SensorGuard-RV on an Artix-7 FPGA with a NEORV32 core and an ADXL362 SPI accelerometer. At 100\,MHz, a protected sample reaches a register in 5,191 cycles (51.9\,$\mu$s), sustaining 19,015 transactions/s (5.3\% above a software-driven SPI baseline). The added logic (3,155 LUTs and 2,215 registers; 5.0\% and 1.7\% of the FPGA) is dominated by the Ascon engine and preserves the 100\,MHz clock, while the estimated power consumption rises from 118 to 368\,mW.
\end{abstract}




\maketitle

\section{Introduction}

Sensor data is critical to the correct and safe operation of cyber-physical systems (CPS), from stabilising unmanned vehicles (UxVs) using accelerometers and gyroscopes, to controlling industrial processes using environmental measurements. On the device that takes those measurements, data must traverse a complex pathway involving peripheral interfaces, communication buses and memory. Unauthorised access to, or mis-configuring, this path may allow an adversary to intercept sensitive measurements, substitute fabricated values or replay earlier samples. A recent case with Bosch's SensorAPI showed how a malformed response from a spoofed or compromised sensor over I$^2$C or serial peripheral interface (SPI) could be used to read and corrupt measurement fields and stack memory (CVE-2026-42806)~\cite{bosch2026sensorapi}. Chen et al.~\cite{chen2023infinitygauntlet} illustrated how fingerprint authentication systems could be compromised by manipulating SPI transactions to spoof fingerprint attempts beyond the rate limits for unsuccessful attempts. Moving up the stack, SMASheD~\cite{mohamed2016smashed} demonstrated how a helper service launched through Android's debugging interface could enable malicious applications to read and inject sensor events consumed by other applications. Beyond manipulating measurements directly, the configuration of sensors provides additional attack vectors, enabling adversaries to mis-calibrate sensors, profile other measurement-consuming applications, and change how readings are produced or interpreted~\cite{shepherd2023side,erba2025confusense,quarta2017robot,mitreCWE1314}. For example, ConfuSense~\cite{erba2025confusense} showed that injected bus commands could change drone sensors' operating modes and sampling rates, disrupting the measurements available to the flight controller.

Attempts to secure the configuration, acquisition and delivery of sensor measurements have been achieved by targeting different parts of this pathway using distinct approaches. This has involved the application of trusted execution environments (TEEs)~\cite{ref07}, trusted platform modules (TPMs)~\cite{ref01,ref02,ref03}, verifiable trusted paths~\cite{ref04}, and hardware-enforced restrictions imposed on sensing software~\cite{ref08}. These mechanisms depend on the software and hardware that maintain the protected path, which can be relatively complicated. In systems that carry plaintext measurements through access-controlled interconnects and memory, security depends on correctly enforcing those access restrictions throughout the transfer, which exposes additional vulnerabilities. For example, a vulnerability (CVE-2016-10423) allowed a trusted application in Qualcomm's TEE to read data from an SPI interface opened by another trusted application because access was not exclusive~\cite{cerdeira2020trustzone}, while insecure controller settings allowed normal-world software to bypass Arm TrustZone's memory protection mechanisms on NXP i.MX platforms (CVE-2021-36133)~\cite{optee2021csu}.

In this work, we present \schemename{}, which tightly couples the sensor acquisition and delivery pathway to hardware-level cryptographic protection. Following initial configuration, we develop a write-once ownership switch to give a dedicated controller exclusive access to the sensor, thus preventing software from reading raw responses or changing its configuration until reset. Each acquired measurement is subsequently encrypted and authenticated prior to transportation over hardware interconnects and memory. At the receiving end, we implement a new RISC-V processor instruction that completes authenticated decryption and releases the plaintext into a register only after verifying the packet and rejecting replay. Keys remain within hardware, and a per-boot nonce epoch prevents packets recorded before a reset from being accepted afterwards. Our proposed design places acquisition and the decision to release a verified measurement in hardware, allowing the intervening software and buffers to remain untrusted; the protection boundary of \schemename{} extends from controller acquisition to processor-register delivery. Within this boundary, an adversary may observe, alter or withhold encrypted packets, but cannot thereby recover their plaintext or cause a forged measurement to be accepted. We make the following contributions:
\begin{itemize}[leftmargin=*]
\item A hardware acquisition architecture that combines exclusive sensor ownership with authenticated encryption, removing driver software from the trusted path after configuration.
\item A RISC-V instruction set extension that verifies and decrypts a complete protected measurement, rejects replay and permits register writeback only after successful authentication and freshness verification.
\item An FPGA implementation and evaluation evidencing acquisition latency, sustained throughput, hardware and power costs, and rejection of altered packets and incomplete verification contexts.
\end{itemize}

We implement \schemename{} using the open-source NEORV32 processor~\cite{neorv32} on a Nexys A7-100T FPGA, with an ADXL362 accelerometer~\cite{adxl362} as a widely used proof-of-concept target sensor. At the baseline clock frequency of 100~MHz, acquisition through verified delivery takes 51.91~$\mu$s, and the complete processing loop sustains 19,015 protected transactions/s---5.31\% above the software-driven SPI baseline under the same configuration. The additional hardware requires 3,155 lookup tables and 2,215 registers, while estimated full-system power rises from 118 to 368~mW.  Our work establishes the feasibility and cost of controller-to-core protection for a fixed acquisition workload using a representative modality. 

\section{Related Work}
\label{sec:related-work}

Trusted sensing has been tackled through sensor attestation, protected execution, peripheral access control, and cryptographic hardware. Early work on mobile sensing used trusted computing hardware to let a recipient check the origin of a reading. Gilbert et al.\ proposed a platform that uses a trusted platform module (TPM) to attest the software responsible for collecting and processing sensor data~\cite{ref01}; Saroiu and Wolman examined two TPM-based sensor designs with different hardware requirements and security guarantees~\cite{ref02}. Liu et al.\ subsequently implemented sensor attestation that includes the code and sensor configuration responsible for a reading. Their sensor-sealing interface also allows an application to recover an encrypted secret only when sensor readings satisfy a specified policy~\cite{ref03}. When a recipient asks the device to compute on its inputs, it must also establish that the reported result came from the requested program. APEX provides this check on low-end microcontrollers by producing a proof of execution and authenticating the output in the presence of compromised software~\cite{ref05}.

Several systems protect the software that acquires or processes measurements. Plug-n-Trust performs medical-data processing in a removable smart card~\cite{ref10}, and Janjua et al.\ use TrustZone to protect acquisition and subsequent transformations~\cite{ref16}. TruSense signs measurements within a trusted sensing framework so that a remote recipient can verify them~\cite{trusense}. Where data must pass between devices, Shepherd et al.\ establish mutually attested channels between trusted execution environments (TEEs)~\cite{ref06}, and Segers et al.\ protect environmental measurements on their path from the sensing device to a server~\cite{ref17}. Noorman et al.\ build on the Sancus protected-module architecture to secure control of input and output devices and to authenticate events exchanged between modules~\cite{noorman2017authentic}. These designs retain software within the trusted part of the sensing system, allowing it to process measurements before they are released or transmitted.

Protecting that software also requires controlling the peripherals from which it obtains its inputs. Zhou et al.\ use a hypervisor to establish a trusted path on x86~\cite{ref04}, and SeCloak uses TrustZone to enforce peripheral enablement policies despite a compromised operating system~\cite{ref11}. HECTOR-V supports peripheral ownership in a RISC-V system through tagged bus transactions and a security monitor that handles claiming, release, and withdrawal; immutable bindings are also available~\cite{ref12}. CURE similarly assigns peripherals exclusively to individual enclaves on a RISC-V system-on-chip~\cite{cure}. VERSA restricts sensing to authorised routines and uses a formally verified hardware monitor to enforce atomic execution and cleanup~\cite{ref08}. Fortress places peripheral memory and selected driver operations within TrustZone, allowing trusted application code to filter or encrypt the acquired data~\cite{ref07}. SGXIO uses trusted drivers with exclusive device access and authenticated encrypted channels to application enclaves, so the intervening operating system need not be trusted to carry their traffic~\cite{sgxio}.

Cryptographic hardware can also protect a reading before ordinary software receives it and reduce CPU verification time. Mart\'{i}nez-Rodr\'{i}guez et al.\ evaluate HMAC, AEGIS-128, and Ascon-128 modules for trusted virtual sensors with keys derived from SRAM physical unclonable functions (PUFs)~\cite{ref09}; Sampaio et al.\ propose TPM/PUF gatekeepers for sensor acquisition~\cite{ref18}. At the processor, Alt{\i}nay and \"{O}rs add Ascon-specific RV32I instructions and compiler support~\cite{ref14}. Steinegger and Primas connect an Ascon permutation accelerator to the register file~\cite{ref15}, and Cheng et al.\ evaluate cryptographic extensions that work within conventional operand limits without adding execution state~\cite{ref13}. Nevertheless, software remains responsible for verifying results before using the plaintext.

For our repeated register-read workload, we place acquisition and packet verification in fixed hardware units, with software copying the encrypted packet between them. Table~\ref{tab:related} compares the resulting protection with the systems discussed above. The access-control designs keep trusted software on the acquisition path; \schemename{} replaces it with a fixed-function controller after configuration, and its one-time ownership transfer fixes the permitted transaction until reset. Encrypting at the controller allows the packet to pass through untrusted bus and memory interfaces, as SGXIO's encrypted channel allows traffic to pass through an untrusted operating system. At the receiving end, our proposed Custom Functions Unit (CFU) completes authenticated decryption and permits plaintext writeback only after tag verification, removing the release decision from the transport software. This arrangement requires a sensor-specific acquisition front end and serialised access to the CFU's staged tag context. Unlike a TEE-based design, \schemename{} does not isolate or attest the consuming program, which relies on privilege separation.

A simpler alternative on NEORV32 would retain a trusted machine-mode driver and use physical memory protection (PMP) to restrict the SPI peripheral and the sample buffer to it. That design needs no cryptography, but it keeps driver software in the trusted computing base, depends on correct PMP configuration for every region that the sample traverses, and does not constrain DMA masters, whose accesses PMP does not check. \schemename{} still relies on PMP to protect the consumer's own state, but not the acquisition path: it removes the driver after handover and makes each packet independently verifiable, so neither the buffer nor the code that copies it needs to be trusted.

\begin{table*}[t]
\caption{Comparison with related work.}
\label{tab:related}
\centering
\footnotesize
\setlength{\tabcolsep}{2pt}
\renewcommand{\arraystretch}{1.08}
\renewcommand{\tabularxcolumn}[1]{m{#1}}
\providecommand{\cmark}{\ding{51}}
\providecommand{\pmark}{\ensuremath{\triangle}}
\providecommand{\xmark}{\textcolor{black!50}{\ensuremath{\times}}}
\providecommand{\rwhead}[1]{\textbf{\shortstack{#1}}}
\begin{tabularx}{\textwidth}{@{}
  >{\raggedright\arraybackslash}m{23mm}
  >{\raggedright\arraybackslash}X
  >{\centering\arraybackslash}m{12mm}
  >{\centering\arraybackslash}m{12mm}
  >{\centering\arraybackslash}m{11mm}
  >{\centering\arraybackslash}m{8mm}
  >{\centering\arraybackslash}m{8mm}
  >{\centering\arraybackslash}m{12mm}
  >{\centering\arraybackslash}m{11mm}
  >{\raggedright\arraybackslash}m{15mm}@{}}
\toprule
\rowcolor{black!7}
\textbf{Scheme} & \textbf{Protection approach} &
\rwhead{Plaintext\\on bus} & \rwhead{Exclusive\\access} &
\rwhead{Source\\protect.} & \rwhead{Conf.} & \rwhead{Integ.} &
\rwhead{Protected\\buffer} & \rwhead{Auth.\\release} & \textbf{Evaluation} \\
\midrule
\rowcolor{black!4}
\multicolumn{10}{@{}l@{}}{\strut\textit{TPM-based trusted sensing and provenance}} \\
Gilbert et al.~\cite{ref01} & TPM-attested trusted domain for sensor acquisition and processing & N/S & \pmark & \pmark & \pmark & \cmark & \cmark & \xmark & Architecture \\
Saroiu and Wolman~\cite{ref02} & TPM-signed readings via an isolated attester or sensor-integrated TPM & Yes & \pmark & \pmark & \xmark & \cmark & \cmark & \xmark & Concept \\
Liu et al.~\cite{ref03} & Sensor attestation and sealing in isolated x86/TrustZone environments & Yes & \cmark & \pmark & \pmark & \cmark & \cmark & \pmark & Prototype \\
\addlinespace[2pt]
\rowcolor{black!4}
\multicolumn{10}{@{}l@{}}{\strut\textit{TEE- and secure-element-based sensing}} \\
Plug-n-Trust~\cite{ref10} & Smart-card processing of encrypted sensor data & No & \pmark & \cmark & \cmark & \cmark & \cmark & \xmark & Prototype \\
Janjua et al.~\cite{ref16} & TrustZone acquisition and trusted data transformations & Varies & \pmark & \pmark & \cmark & \cmark & \pmark & \xmark & Prototype \\
TruSense~\cite{trusense} & TrustZone measurement signing and boot attestation & Yes & \cmark & \pmark & \xmark & \cmark & \cmark & \xmark & Prototype \\
Shepherd et al.~\cite{ref06} & Mutually attested, encrypted channels between TEEs & N/S & \xmark & \pmark & \cmark & \cmark & \cmark & \xmark & Prototype \\
Segers et al.~\cite{ref17} & TrustZone sensing and authenticated encryption to a server & Yes & \pmark & \pmark & \cmark & \cmark & \cmark & \xmark & Prototype \\
\addlinespace[2pt]
\rowcolor{black!4}
\multicolumn{10}{@{}l@{}}{\strut\textit{Trusted I/O and hardware monitors}} \\
SGXIO~\cite{sgxio} & Exclusive device binding and encrypted driver--enclave channels & Yes & \cmark & \cmark & \cmark & \cmark & \cmark & \pmark & Prototype \\
Fortress~\cite{ref07} & TrustZone isolation of peripheral I/O and trusted drivers & Yes & \cmark & \pmark & \cmark & \cmark & \pmark & \xmark & Prototype \\
VERSA~\cite{ref08} & Verified hardware enforcement of authorised sensing routines & Yes & \cmark & \pmark & \cmark & \pmark & \pmark & \xmark & Formal; FPGA \\
HECTOR-V~\cite{ref12} & Tagged-bus access control and peripheral ownership & Yes & \cmark & \xmark & \cmark & \cmark & \xmark & \xmark & FPGA \\
\addlinespace[2pt]
\rowcolor{black!4}
\multicolumn{10}{@{}l@{}}{\strut\textit{Sensor-side cryptographic protection}} \\
Virtual sensors~\cite{ref09} & PUF-derived keys for sensor-output MAC or AEAD & Varies & \xmark & \cmark & \cmark & \cmark & \cmark & \xmark & VLSI eval. \\
Sampaio et al.~\cite{ref18} & Proposed TPM/PUF authentication of sensors and microcontrollers & N/S & \pmark & \pmark & \xmark & \pmark & \pmark & \xmark & Concept \\
\midrule
\rowcolor{blue!5}
\textbf{\schemename{}} & Exclusive SPI acquisition, AEAD, and CFU verification before release & No & \cmark & \cmark & \cmark & \cmark & \cmark & \cmark & \textbf{FPGA} \\
\bottomrule
\end{tabularx}
\par\smallskip
\begin{minipage}{\textwidth}
\scriptsize\raggedright
\textbf{Legend:} \cmark{} provided at the stated boundary;
\pmark{} partial, conditional, or proposed;
\xmark{} not claimed.
\textbf{Plaintext on bus:} whether data traverses the host's system-I/O bus before processing; N/S = not specified for acquisition.
\textbf{Other columns:} exclusive access = hardware restriction of sensor access;
source protect.\ = protection applied before exposure to an untrusted path;
conf./integ. = confidentiality/integrity within the stated boundary, through isolation or cryptography;
protected buffer = protected output may traverse untrusted memory or communication (\pmark{} = optional or application-dependent); auth.\ release = local release after packet verification (\pmark{} = protected-software verification rather than a dedicated hardware gate).
For the TPM rows, integrity and protected-buffer support refer to attested outputs, not ciphertext.
\end{minipage}
\end{table*}

\section{Hardware-Software Design}
\label{sec:architecture}


Figure~\ref{fig:sable-sense-arch} shows how a sample passes from the trusted acquisition controller to the CFU through ordinary bus and memory interfaces. The acquisition and protection wrapper (HW1) contains the SPI controller, sample-packing logic, an Ascon-128a encryptor, and packet-publication registers. Before this wrapper can acquire data, the ownership switch (HW2) must select its SPI master and disconnect the standard peripheral's receive path. HW1 can then obtain and protect a sample without exposing plaintext to software, which reads only the published packet and presents it to the CFU. The CFU performs authenticated decryption and supplies the result to the processor register file through the custom-instruction.

Each packet comprises $P=(s,C,T)$, with a 32-bit sequence number, $s$, and ciphertext, $C$, and a 128-bit tag $T$. The sequence number supplies the per-packet part of the nonce. With a 96-bit per-boot epoch ($E$) held in hardware, it allows the encryptor and decryptor to derive the same nonce without transmitting all 128 bits. The ciphertext occupies one processor operand, while the tag is loaded separately into the CFU before decryption. Authenticating each acquisition individually allows the controller to publish it as soon as encryption finishes, without waiting for a batch of samples. It also expands a four-byte measurement into a 24-byte packet, whose six-word transfer through memory is included in the measured transaction cost. The ownership and verification mechanisms could also serve other sensors, with changes to the acquisition front end, packet format, and processor interface as discussed in \S\ref{sec:limitations}.

\begin{figure}[t]
\centering
\resizebox{\columnwidth}{!}{
\definecolor{ssgreen}{RGB}{20,75,35}
\definecolor{ssred}{RGB}{225,20,20}
\providecommand{\circled}[1]{\tikz[baseline=(c.base)]\node[circle,draw,inner sep=0.4pt,font=\sffamily\tiny\bfseries](c){#1};}
\begin{tikzpicture}[
  font=\sffamily\footnotesize,
  >={Stealth[length=1.8mm]},
  gb/.style={draw=ssgreen, thick, rounded corners=2pt, fill=white,
             align=center, minimum height=0.8cm, text=black,
             inner sep=2pt},
  bb/.style={draw=ssgreen, thick, rounded corners=2pt, fill=white,
             align=center, minimum height=0.8cm, text=ssgreen,
             inner sep=2pt},
  ub/.style={draw=ssred, thick, rounded corners=2pt, fill=white,
             align=center, minimum height=0.8cm, text=black,
             inner sep=2pt},
  sg/.style={fill=ssgreen!14},
  ar/.style={->, semithick},
  lb/.style={font=\sffamily\scriptsize, align=center},
  noentry/.pic={
    \draw[ssred, thick, fill=white] (0,0) circle (0.13);
    \draw[ssred, thick] (-0.09,0.09) -- (0.09,-0.09);},
  lock/.pic={
    \draw[ssgreen, line width=1.1pt] (-0.06,0.02) -- (-0.06,0.09)
        arc (180:0:0.06) -- (0.06,0.02);
    \fill[ssgreen] (-0.11,-0.15) rectangle (0.11,0.03);},
  shield/.pic={
    \fill[ssgreen] (0,0.28) .. controls (0.13,0.22) .. (0.24,0.22)
        -- (0.24,0.01) .. controls (0.22,-0.16) .. (0,-0.30)
        .. controls (-0.22,-0.16) .. (-0.24,0.01)
        -- (-0.24,0.22) .. controls (-0.13,0.22) .. (0,0.28);
    \draw[white, line width=1.2pt, line cap=round, line join=round]
        (-0.10,0.0) -- (-0.02,-0.09) -- (0.12,0.10);}
]

\node[gb, minimum width=2.9cm, minimum height=1.0cm] (adxl) at (0,0.5)
  {\textbf{ADXL362 Accelerometer}\\\scriptsize Trusted physical sensor};

\draw[ssgreen, thick, fill=ssgreen!14, rounded corners=1.5pt]
  (-1.1,-0.9) -- (1.1,-0.9) -- (0.7,-1.8) -- (-0.7,-1.8) -- cycle;
\node[align=center, font=\sffamily\scriptsize] at (0,-1.33)
  {SPI Ownership\\Switch (HW2)};

\node[gb, minimum width=2.6cm] (spi) at (4.5,-1.05)
  {\textbf{NEORV32 SPI}\\\scriptsize Initialisation only};

\draw[ar] (adxl.south) -- (0,-0.9)
  node[midway, left, lb, align=right] {Protected SPI:\\MOSI, MISO,\\CSN, SCLK};
\draw[ar] (spi.west) -- (1.033,-1.05);
\node[lb, text=ssred, anchor=north] at (2.15,-1.12) {MISO $\to$ 0};

\draw[ssred, semithick, dashed, ->] (spi.north) |- (adxl.east);
\pic at (3.0,0.5) {noentry};
\pic at (4.5,-0.1) {noentry};
\node[lb, text=ssred, anchor=south] at (4.6,0.66)
  {CPU raw-data path\\blocked after handover};

\node[gb, sg, minimum width=2.9cm] (acq) at (0,-2.75)
  {\textbf{Fixed-Function Acquisition}\\\scriptsize \texttt{0x0B} burst read};
\node[gb, sg, minimum width=2.9cm] (pack) at (0,-3.85)
  {\textbf{Packed Sample}\\\scriptsize Y[15:0] $\|$ X[15:0]};
\node[gb, sg, minimum width=2.9cm] (enc) at (0,-4.95)
  {\textbf{Ascon-128a Encryptor}};
\node[gb, sg, minimum width=3.3cm, minimum height=1.2cm] (pkt) at (0,-6.25)
  {\textbf{\schemename{} Packet Registers}\\[1pt]
   \fbox{\scriptsize $s$ (32) $|$ $C$ (32) $|$ $T$ (128)}\\[1pt]
   \scriptsize VALID / ACK};

\draw[ar] (acq.north)  -- (0,-1.8);
\draw[ar] (acq)  -- (pack);
\draw[ar] (pack) -- (enc);
\draw[ar] (enc)  -- (pkt);

\pic at (2.15,-4.95) {lock};
\node[lb, anchor=west, align=left] (key) at (2.35,-4.95) {Key, nonce,\\TRNG epoch};
\draw[ar] (1.95,-4.95) -- (enc.east);

\begin{scope}[on background layer]
  \node[draw=ssgreen, dashed, rounded corners=4pt,
        fit={(-1.95,-2.05) (4.1,-7.1)}, inner sep=0pt] (prot) {};
  \node[font=\sffamily\scriptsize\bfseries, text=ssgreen, anchor=north east]
    at (4.0,-2.12) {HW1};
  \node[draw=ssgreen, line width=1pt, rounded corners=5pt,
        fit={(-2.3,2.0) (7.0,-7.4)}, inner sep=0pt] (zone) {};
\end{scope}
\pic at (-1.95,-6.75) {shield};
\node[font=\sffamily\bfseries\small, text=ssgreen] at (2.35,1.65)
  {Trusted Components};

\node[draw=ssred, thick, rounded corners=2pt, fill=white, text=ssred,
      minimum width=9.3cm, minimum height=0.5cm, inner sep=1pt]
  (bus) at (2.35,-8.4) {NEORV32 System Bus (untrusted)};

\draw[ar] (6.6,-8.15) -- (6.6,-1.05)
  node[pos=0.45, sloped, above, lb] {Initialisation commands} -- (spi.east);
\draw[ar] (6.2,-8.15) -- (6.2,-1.65)
  node[pos=0.45, sloped, above, lb] {TAKE\_OWNERSHIP} -- (0.767,-1.65);
\draw[ar] (5.8,-8.15) -- (5.8,-6.25)
  node[pos=0.5, sloped, above, lb] {START / ACK} -- (pkt.east);

\node[draw=ssred, thick, rounded corners=2pt, fill=white, align=center,
      minimum width=2.6cm, minimum height=0.75cm, inner sep=1pt]
  (dmem) at (5.4,-9.45)
  {\textbf{Data Memory (DMEM)}\\\scriptsize holds only $(s,C,T)$};
\draw[<->, semithick] (5.4,-8.65) -- (dmem.north);
\begin{scope}[on background layer]
  \node[draw=ssred, dashed, thick, rounded corners=4pt,
        fit={(-2.3,-8.08) (7.0,-10.05)}, inner sep=0pt] (untr) {};
\end{scope}
\node[lb, text=ssred, anchor=south west, align=left] at (-2.2,-10.0)
  {\textbf{Untrusted}};

\node[ub, minimum width=2.0cm] (poll) at (0,-11.25) {\textbf{Polling}\\\textbf{Load/Store}};
\node[font=\sffamily\bfseries\scriptsize, text=ssred, anchor=south west,
      align=left] at (-2.15,-12.68) {NEORV32 software\\(untrusted)};
\begin{scope}[on background layer]
  \node[draw=ssred, dashed, thick, rounded corners=5pt,
        fit={(-2.3,-10.4) (1.15,-12.75)}, inner sep=0pt] (sw) {};
\end{scope}

\node[bb, sg, text=black, minimum width=2.2cm] (cfu)  at (2.65,-11.25)
  {\textbf{Ascon CFU}\\Decrypt + Verify};
\node[bb, text=black, minimum width=1.7cm] (rf)   at (5.15,-11.25)
  {\textbf{Register}\\\textbf{File}};
\draw[ar, very thick, ssgreen] (cfu.east) -- (rf.west)
  node[midway, above, lb, text=ssgreen, yshift=1pt] {\circled{4}};
\node[lb, text=black, align=center] at (4.2,-12.02)
  {M-mode only; X/Y written only if the tag\\verifies and the sequence is fresh};
\node[font=\sffamily\bfseries\footnotesize, text=ssgreen, anchor=south west]
  at (1.45,-12.68) {NEORV32 CFU (trusted)};
\begin{scope}[on background layer]
  \node[draw=ssgreen, line width=1pt, rounded corners=5pt,
        fit={(1.3,-10.4) (7.0,-12.75)}, inner sep=0pt] (cpu) {};
\end{scope}

\draw[ar] ($(pkt.south)+(-0.6,0)$) -- (-0.6,-8.15);
\draw[ar] ($(pkt.south)+(0.6,0)$)  -- (0.6,-8.15);
\node[lb, anchor=east] at (-0.65,-7.8) {Packet\\$(s,C,T)$};
\node[lb, anchor=west, align=left] at (0.65,-7.8)
  {STATUS (VALID, BUSY,\\OVERRUN, OWNED)};

\draw[ar] (-0.6,-8.65) -- (-0.6,-10.85);
\node[lb, anchor=east] at (-0.65,-9.05) {\circled{1} read};
\draw[ar] (0.6,-10.85) -- (0.6,-9.45)
  -- node[above, lb] {\circled{2} store $(s,C,T)$} (dmem.west);
\draw[ar] (4.6,-9.83) -- (4.6,-10.62) -| (3.2,-10.85);
\node[lb, anchor=west] at (4.65,-10.22) {\circled{3} load $(s,C,T)$};
\end{tikzpicture}}
\caption{\schemename{} architecture.}
\Description{A vertical path from the trusted ADXL362 sensor through the write-once ownership switch and the acquisition and encryption wrapper to the packet registers. The standard SPI peripheral's raw-data path is blocked after handover. An untrusted system bus, data memory, and untrusted processor software carry the encrypted packet to the trusted Ascon CFU, which writes the decrypted sample to the register file only after tag verification and a sequence check.}
\label{fig:sable-sense-arch}
\vspace{-0.2cm}
\end{figure}

\subsection{Trust Model and Security Goals}
\label{sec:threat-model}

We now describe the trust and security assumptions that are relied upon in our proposed system.

\emph{Trusted computing base.} We trust the ADXL362 and its board-level SPI connection; the ownership switch, acquisition controller, both Ascon engines, and the CFU's execution and writeback logic; the FPGA configuration; the hardware-provisioned key and epoch ($E$), which have no software-visible read or write interface; and the firmware that configures the sensor before handover. Protection begins when ownership transfers to the acquisition controller and ends when a verified sample reaches the destination register. Computation on the released sample, register spills, and stores of it fall outside this interval.

\emph{Adversary.} The adversary controls software running below machine mode, together with any bus master that can reach data memory or memory-mapped I/O. It can observe, modify, replay, reorder, drop, or mix packet fields on the bus and in memory; issue any wrapper command, including \texttt{START}, \texttt{ACK}, and \texttt{TAKE\_OWNERSHIP}; drive the standard SPI peripheral; attempt to invoke the CFU with arbitrary operands; and reset the system.

\emph{Consumer.} The consumer is the machine-mode code that invokes \texttt{AUDEC} and acts on its result. We assume that it executes as written and that PMP keeps its state, including released samples, from less-privileged software, with an I/O PMP providing the same protection against DMA masters. Without this assumption, no local mechanism can protect a reading, since the adversary could overwrite whatever value the consumer uses. The evaluated bare-metal program runs entirely in machine mode; its packet copies through ordinary memory model the untrusted transport path.

\emph{Goals.} (G1)~\emph{Exclusive control}: after handover, no software can read raw sensor responses or issue sensor transactions. (G2)~\emph{Authenticity}: \texttt{AUDEC} sets \texttt{AUTH\_OK} only for a sample that the controller produced. (G3)~\emph{Confidentiality}: a packet in transit reveals nothing about its sample to the adversary. (G4)~\emph{Replay rejection}: \texttt{AUDEC} accepts packets only in increasing sequence order within a boot and rejects packets produced before the most recent reset. Sequence order establishes the order of acquisitions, not their age.


\emph{Out of scope.} Machine-mode firmware is trusted, which can be achieved using a secure boot process as standard in related systems ~\cite{cure,keystone,ref12}. The untrusted region denotes the packet transport path, whose contents are verified before plaintext release. Physical attacks on the sensor or board, compromised FPGA configurations, side channels, fault injection, key extraction, and manipulation of the measured physical quantity are out of scope. The adversary can deny service by withholding requests, acknowledging packets prematurely, or resetting the system, but cannot thereby forge or replay a sample. Section~\ref{sec:security-analysis} maps each adversary capability to the mechanism that addresses it.

\subsection{Sensor Ownership and Acquisition}

Trusted firmware configures the sensor once, after which every acquisition repeats the same register read. The processor therefore no longer needs to drive the SPI pins, and a write-once switch can give HW1 exclusive ownership until reset. Software can still request measurements, but it cannot change the acquisition transaction or reconfigure the sensor. Supporting either operation at runtime would require a more flexible access policy, such as HECTOR-V's peripheral claiming and release~\cite{ref12} or VERSA's authorisation of sensing routines~\cite{ref08}. To enforce exclusive ownership in our design, HW2 gates the receive signal as well as selecting the master. Without this gate, software may continue reading raw responses through the peripheral despite HW1 encrypting the same measurements.

After reset, the standard NEORV32 SPI peripheral drives the serial clock (SCLK), master-out/slave-in (MOSI), and active-low chip-select (CSN) signals and receives physical MISO. Once configuration is complete, trusted firmware waits for the interface and transmit first-in-first-out (FIFO) buffer to become idle before issuing \texttt{TAKE\_OWNERSHIP}. HW2 accepts the command only when both masters present the mode-0 idle levels, CSN high and SCLK low, and the acquisition controller is not busy. An unsafe attempt produces a one-cycle rejection indication and leaves ownership unchanged. Following acceptance, multiplexers route the trusted controller's outputs to the sensor, and symmetric MISO gating delivers responses only to that controller while forcing the non-owner's input to zero. Ownership remains latched until hardware reset, preventing the standard peripheral from either initiating another sensor transaction or observing its response.

With ownership established, HW1 accepts \texttt{START} only while the acquisition and encryption pipeline is idle and no earlier packet awaits acknowledgement. The controller then performs a six-byte mode-0 burst, transmitting the ADXL362 read command \texttt{0x0B}, starting address \texttt{0x0E}, and four dummy bytes while receiving \texttt{X\_L}, \texttt{X\_H}, \texttt{Y\_L}, and \texttt{Y\_H}. Each axis is represented as a signed 12-bit value~\cite{adxl362}. The controller combines the low byte with the lower nibble of the high byte, sign-extends the result to 16 bits, and packs the pair as \(M=Y[15\!:\!0]\mathbin{\|}X[15\!:\!0]\), where \(\mathbin{\|}\) denotes concatenation. Keeping this X/Y payload to 32 bits allows both its ciphertext and the recovered sample to fit individual processor registers. Supporting the Z axis or temperature channel would require enlarging the payload and adapting the processor interface.

\subsection{Encryption and Packet Publication}
\label{sec:encryption}

If a sample reaches ordinary memory before authentication, an attacker can replace it before software computes its tag. HW1 therefore encrypts and authenticates the sample during acquisition, following the source-side protection evaluated for trusted virtual sensors by Mart\'{i}nez-Rodr\'{i}guez et al.~\cite{ref09}. Software receives a complete packet that the CFU can verify without trusting the intervening transfers. We authenticate each sample separately so that HW1 can publish it as soon as encryption finishes, at the cost of carrying a sequence number and tag with every four-byte measurement. The ownership switch makes this the only software-accessible acquisition path after handover.

When ownership is transferred, hardware draws a 96-bit random epoch $E$ from NEORV32's TRNG and latches it for both the encryptor and the CFU, with no software-visible read or write interface. When HW1 accepts \texttt{START}, it captures the current 32-bit sequence number $s$ and derives the nonce ($N$) with \(N[31\!:\!0]= s\) and \(N[127\!:\!32]= E\) so that \(N=E\mathbin{\|}s\). The CFU applies the same derivation using its internal copy of the 128-bit key and 96-bit \(E\), none of which is exposed through memory-mapped registers or CPU operands. The transmitted sequence number therefore suffices to reconstruct the nonce within a boot. The evaluated format uses a four-byte plaintext and zero-length associated data. Sensor identity and configuration are fixed deployment parameters; they are not carried as authenticated packet fields.

The engine implements Ascon-128a~\cite{ascon-v12}, conforming to the NIST ASCON standard~\cite{nist-ascon}. With empty associated data (AD), a four-byte message fits within a single rate block, requiring only the two Ascon-$p[12]$ permutations used for initialisation and finalisation. Between these permutations, the engine handles domain separation for the empty AD, plaintext processing, and padding prior to tag generation. Each 12-round permutation executes as six two-round passes through shared logic, enabling hardware reuse while meeting the 100-MHz performance target. The evaluation in \S\ref{sec:ascon-dse} compares this architecture against alternative folding factors, measuring the latency of the complete authenticated operation, including the initialisation and finalisation overhead required for every sample.

Once encryption finishes, HW1 copies $(s,C,T)$ into \texttt{SEQUENCE}, \texttt{CIPHERTEXT}, and \texttt{TAG0}--\texttt{TAG3} (six 32-bit memory-mapped registers), asserting \texttt{VALID} on the same edge. All six words remain stable until software issues \texttt{ACK}, so a reader can finish copying the packet before another acquisition replaces it. The \texttt{CTRL} register provides the one-shot \texttt{ACK}, \texttt{CLR\_OVERRUN}, \texttt{START}, and \texttt{TAKE\_OWNERSHIP} commands, while \texttt{STATUS} reports \texttt{VALID}, \texttt{BUSY}, \texttt{OVERRUN}, and \texttt{OWNED}. A request made without ownership, while busy, or while an unread packet remains valid is rejected and sets sticky \texttt{OVERRUN}. Command acceptance uses the state at the beginning of the cycle, which means that a combined \texttt{ACK}/\texttt{START} write acknowledges the current packet but cannot launch the next acquisition. The handshake prevents a reader following the protocol from mixing packet generations; fields mixed from different packets fail CFU authentication.

\subsection{Authenticated-Decryption CFU}
\label{sec:cfu}

Accelerating the Ascon permutation, as in the extensions of Steinegger and Primas~\cite{ref15} and Cheng et al.~\cite{ref13}, leaves software to complete authenticated decryption and decide whether its result can be used. Our instruction performs the complete operation under a hardware-only key, so that the tag comparison directly controls architectural writeback. A software routine built from such extensions must keep the key in software-visible state and produces the plaintext before checking the tag; an adversary able to read or write data memory could then recover the key or replace the plaintext between the check and its use. The 32-bit sequence and ciphertext fit the two source registers, and the recovered sample fits the destination register. Supplying the 128-bit tag requires four preceding writes to an internal CFU context, whose validity mask records whether all tag words have arrived. This departs from the stateless instruction approach of Cheng et al.: callers must serialise access to the shared context because the prototype provides no per-task state.

The CFU connects to NEORV32 through the native custom-instruction start, decode, operand, result, and completion signals. Four single-cycle I-type operations load the tag, each copying \texttt{rs1} into one of \texttt{TAG0}--\texttt{TAG3} and setting the corresponding bit in a four-bit validity mask. A separate single-cycle status-read operation exposes persistent \texttt{AUTH\_OK}, \texttt{DONE}, \texttt{BUSY}, and \texttt{CONTEXT\_ERROR} in bits 0--3. Once software has supplied the tag, the R-type \texttt{AUDEC} instruction takes the sequence in \texttt{rs1} and ciphertext in \texttt{rs2} and waits for authenticated decryption to complete. The CFU executes \texttt{AUDEC} only in machine mode and rejects invocations from less-privileged modes, so user-mode software cannot decrypt a packet even if it holds a valid one. If a tag word is missing, the CFU instead returns zero immediately with \texttt{DONE} and \texttt{CONTEXT\_ERROR}, without starting the Ascon engine. Every attempt consumes the validity mask, so a later invocation cannot accidentally reuse the previous tag context. With all words present, the CFU assembles \(T=\texttt{TAG3}\mathbin{\|}\texttt{TAG2}\mathbin{\|}\texttt{TAG1}\mathbin{\|}\texttt{TAG0}\), reconstructs the nonce, and decrypts using the same six two-round passes per permutation as the source encryptor.

The recovered plaintext is withheld from architectural writeback while a fixed-latency comparison checks the supplied and reconstructed tags. The CFU also retains the sequence number of the last accepted packet and accepts a new packet only if its sequence number is greater. Because the sequence number determines the nonce, an adversary cannot raise it without invalidating the tag. A match with a fresh sequence number permits writeback and asserts \texttt{AUTH\_OK}; a tag mismatch returns zero with \texttt{DONE} asserted and \texttt{AUTH\_OK} clear, and a replayed packet is likewise rejected without writeback. In every case, the instruction completes so that the processor can continue and inspect status instead of waiting indefinitely. The consumer must check that status because a genuine all-zero X/Y sample has the same returned value as a failed operation, but it cannot obtain an unverified or replayed sample through the instruction. The fixed-latency comparison does not by itself establish resistance to timing or other side-channel attacks.

\subsection{Software Workflow}

\begin{figure}[h]
\centering
\resizebox{\columnwidth}{!}{%
\begin{tikzpicture}[
  node distance=2.3mm,
  font=\scriptsize,
  process/.style={
    rectangle, rounded corners=1.5pt, draw,
    align=center, text width=61mm,
    minimum height=7mm, inner sep=2pt
  },
  decision/.style={
    diamond, aspect=2.4, draw,
    align=center, inner sep=0.5pt
  },
  failure/.style={
    rectangle, rounded corners=1.5pt, draw,
    align=center, text width=17mm,
    minimum height=9mm, inner sep=2pt
  },
  flow/.style={-{Latex[length=1.7mm]}, semithick},
  line/.style={semithick},
  phase/.style={font=\footnotesize\bfseries, align=center},
  lbl/.style={font=\scriptsize, inner sep=1.5pt}
]

\node[phase] (phase1) {One-time initialisation and handover};

\node[process, below=1.5mm of phase1] (config) {
  Initialise NEORV32; configure SPI and ADXL362;
  reset sensor, verify IDs, and enable
  $\pm 2g$/100-Hz measurement mode
};

\node[process, below=of config] (handover) {
  Wait for the SPI interface and FIFO to become idle;
  issue \texttt{TAKE\_OWNERSHIP}
};

\node[decision, below=of handover] (owned) {
  \texttt{OWNED}\\before timeout?
};

\node[process, below=of owned] (stabilise) {
  Wait 100\,ms for sensor stabilisation
};

\node[phase, below=3mm of stabilise] (phase2) {Repeated protected-sample loop};

\node[process, below=1.5mm of phase2] (start) {
  Clear \texttt{OVERRUN} and issue \texttt{START}
};

\node[decision, below=of start] (valid) {
  \texttt{VALID}\\before timeout?
};

\node[process, below=of valid] (read) {
  Read sequence, ciphertext, and \texttt{TAG0}--\texttt{TAG3};\\
  then issue \texttt{ACK}
};

\node[process, below=of read] (audec) {
  Load the tag context and execute\\
  \texttt{AUDEC(sequence,ciphertext)}
};

\node[decision, below=of audec] (auth) {
  Status exactly\\\texttt{AUTH\_OK | DONE}?
};

\node[process, below=of auth] (use) {
  Check sequence continuity; unpack and use X/Y
};

\coordinate (lane) at ($(config.east)+(12.5mm,0)$);
\node[failure] (fail) at (valid -| lane) {
  Discard result,\\report failure,\\and halt
};

\draw[flow] (config) -- (handover);
\draw[flow] (handover) -- (owned);
\draw[flow] (owned) -- node[lbl, right, pos=0.3] {yes} (stabilise);
\draw[line] (stabilise) -- (phase2);
\draw[flow] (phase2) -- (start);
\draw[flow] (start) -- (valid);
\draw[flow] (valid) -- node[lbl, right, pos=0.3] {yes} (read);
\draw[flow] (read) -- (audec);
\draw[flow] (audec) -- (auth);
\draw[flow] (auth) -- node[lbl, right, pos=0.3] {yes} (use);

\draw[flow] (owned.east) -- node[lbl, above, pos=0.3] {no} (owned.east -| lane)
  -- (fail.north);
\draw[flow] (valid.east) -- node[lbl, above, pos=0.3] {no} (fail.west);
\draw[flow] (auth.east)  -- node[lbl, above, pos=0.3] {no} (auth.east -| lane)
  -- (fail.south);

\coordinate (loop) at ($(use.west)+(-4mm,0)$);
\draw[flow] (use.west) -- (loop)
  -- node[lbl, sloped, above, pos=0.5] {next sample} (loop |- start.west)
  -- (start.west);

\end{tikzpicture}%
}
\caption{Firmware flow. Configuration completes before the irreversible handover; each packet is copied before \texttt{ACK}, and its sample is used only if \texttt{AUDEC} reports \texttt{AUTH\_OK\,|\,DONE}.}
\Description{A flowchart showing one-time ADXL362 initialisation and
irreversible SPI ownership transfer, followed by a repeated loop that
acquires a protected packet, performs authenticated decryption, and
releases X and Y only after successful authentication. Timeout and
authentication failures cause the result to be discarded and execution
to halt.}
\label{fig:software-workflow}
\end{figure}
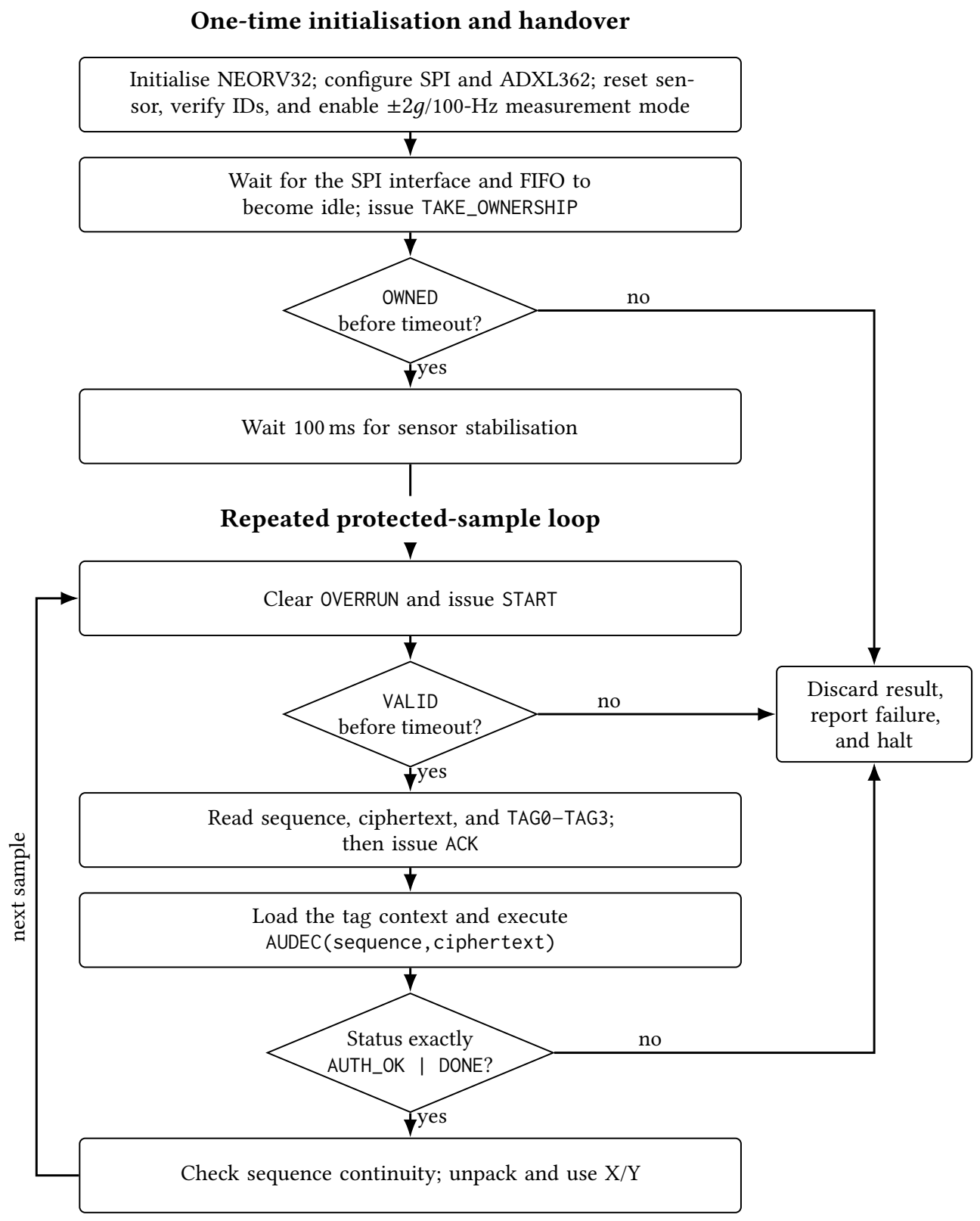

Software completes sensor configuration before transferring ownership, then uses only the packet interface for acquisition, as shown in Figure~\ref{fig:software-workflow}. During initialisation, it sets the standard SPI peripheral to mode~0 at approximately 1~MHz, soft-resets the ADXL362, verifies its identifiers, selects the $\pm2g$ range and 100-Hz output-data rate, and enables measurement mode. Once the SPI interface and transmit FIFO are idle, it clears stale packet state, issues \texttt{TAKE\_OWNERSHIP}, and polls \texttt{OWNED} with a bounded timeout. Handover takes place before a 100-ms sensor-stabilisation delay, shortening the interval during which raw MISO data remain accessible through the standard peripheral. The fixed controller cannot subsequently change this configuration, so all sensor-specific setup must have finished before ownership is transferred.

To acquire a measurement, software clears the prior overrun indication, issues \texttt{START}, and polls \texttt{VALID} with a bounded timeout. It then copies the sequence, ciphertext, and four tag words into memory using ordinary loads and stores, acknowledging the packet only after the copy is complete. After loading the CFU tag context and invoking \texttt{AUDEC(sequence,ciphertext)}, software accepts the returned word only if status equals \texttt{AUTH\_OK | DONE}. It then interprets the lower and upper 16-bit fields as the signed X and Y measurements. The validation program checks sign extension, modified-tag rejection, and incomplete-context rejection before continuous acquisition, and stops on a timeout, sequence discontinuity, overrun, authentication failure, or context error. The continuity check is diagnostic; replay rejection is enforced by the CFU (\S\ref{sec:cfu}). The sequence counts requests rather than distinct physical samples. Access to both the packet registers and CFU is serialised by the evaluated bare-metal program, which runs in machine mode.

\subsection{Security Analysis}
\label{sec:security-analysis}

Table~\ref{tab:security} relates each adversary capability in \S\ref{sec:threat-model} to the mechanism that addresses it. Each mechanism closes a path that the others leave open. Without the receive-path gate, software could read the raw sample through the standard peripheral while the controller encrypted it; without write-once ownership, it could reconfigure the sensor and have the controller authenticate a falsified reading; without source-side encryption, it could substitute the sample in memory; and without a hardware-held key and fused verification, it could recover the key or alter plaintext between a software tag check and its use. The privilege check, sequence check, and per-boot epoch then restrict decryption to the consumer and limit each valid packet to a single acceptance.

\begin{table}[t]
\caption{Adversary capabilities and their outcomes.}
\label{tab:security}
\centering
\footnotesize
\setlength{\tabcolsep}{3pt}
\begin{tabularx}{\columnwidth}{@{}>{\raggedright\arraybackslash}X>{\raggedright\arraybackslash}p{23mm}>{\raggedright\arraybackslash}p{21mm}@{}}
\toprule
Capability & Countermeasure & Outcome \\
\midrule
Read MISO via standard SPI & Receive-path gate & Reads zero (G1) \\[0.1cm]
Reconfigure sensor (range, rate, self-test) & Write-once ownership & No sensor access until reset (G1) \\[0.3cm]
Observe packet on bus or in memory & Ascon engine & Sample encrypted (G3) \\[0.1cm]
Modify $s$, $C$, or $T$; mix packet fields & Tag over $C$ and nonce & Rejected (G2) \\[0.1cm]
Invoke \texttt{AUDEC} below machine mode & Privilege check & Rejected (G3) \\[0.1cm]
Replay a packet in the same boot & Sequence check & Rejected (G4) \\[0.1cm]
Replay a packet from an earlier boot & Per-boot epoch & Rejected (G4) \\[0.1cm]
Supply incomplete tag context & Validity mask & \texttt{CONTEXT\_ERROR} \\
\bottomrule
\end{tabularx}
\end{table}

\section{Experimental Evaluation}
\label{sec:evaluation}

Our prototype extends NEORV32 v1.12.8.3 on a Digilent Nexys A7-100T board with a Xilinx Artix-7 XC7A100T FPGA. Its memory configuration provides 32~KiB for instructions and 64~KiB for data.
To compare protected and conventional acquisition, both builds use a 100~MHz system clock, and access the ADXL362 in SPI mode~0 at approximately 1.04~MHz, with the sensor set to a $\pm2g$ range and 100~Hz output-data rate. Memory capacities and NEORV32 parameters unrelated to the added protection logic are held constant. UART provides host communication, although output is disabled during timed throughput runs. We use Vivado ML Edition~2025.2 for synthesis, placement, and routing, and GCC~15.2.0 from the \texttt{riscv32-unknown-elf} toolchain for firmware compilation.

Latency is measured with a Vivado Integrated Logic Analyzer (ILA), whereas resource, timing, and full-system power results come from separate non-instrumented post-implementation builds. Excluding the ILA from those builds prevents debug logic from affecting the hardware-cost comparison. Power is estimated in Vivado using identical estimation and switching-activity settings for both systems; no board-level power measurements are reported. Our numerical comparison is restricted to the same-platform acquisition baseline because the trusted-I/O systems and cryptographic extensions in \S\ref{sec:related-work} use different hardware, protection boundaries, or message formats.

\subsection{Ascon-128a Engine Selection}
\label{sec:ascon-dse}

We evaluated seven Ascon-128a encryption/decryption implementations to choose an engine that fits the processor's 100-MHz clock and completes authentication with as little instruction delay as possible. Reusing permutation logic reduces area but adds cycles, so the comparison covers both a fully combinational implementation and designs that execute fewer rounds per cycle.
The VHDL modules were developed from C derived from the Ascon Team's official \texttt{ascon-c} implementation~\cite{ascon-c}. The modules were checked against reference-derived ciphertext and tag cases. For the four-byte payload with empty associated data, every operation executes Ascon-$p[12]$ during both initialisation and finalisation. Implementing these permutations combinationally gives a 90-ns critical path, requires 15,603 LUTs, and yields a 456-mW power estimate (at 5 MHz). Its delay exceeds the available clock period, so this design cannot be used in the 100-MHz system.

\begin{figure}[t]
\centering
\definecolor{dseBar}{RGB}{176,176,176}
\definecolor{dseSel}{RGB}{31,95,168}
\begin{tikzpicture}
\begin{groupplot}[
  group style={group size=2 by 2, horizontal sep=9mm, vertical sep=11mm},
  scale only axis, width=31mm, height=21mm,
  ybar, /pgf/bar width=3.1mm, set layers,
  symbolic x coords={12,6,4,3,2,1},
  xtick=data, enlarge x limits=0.13,
  ymin=0,
  tick label style={font=\scriptsize},
  title style={font=\scriptsize\bfseries, yshift=-1.2ex},
  xlabel style={font=\scriptsize},
  axis line style={gray!70}, tick style={gray!70},
  ymajorgrids, grid style={gray!20},
  axis x line*=bottom, axis y line*=left,
  nodes near coords, every node near coord/.append style={font=\tiny, color=black, fill=white, inner sep=0.6pt},
  /pgf/bar shift=0pt,
]

\nextgroupplot[title={(a) Critical path (ns)}, ymax=50, ytick={0,20,30,40,50},
  every node near coord/.append style={yshift=3pt},
  extra y ticks={10}, extra y tick labels={10},
  extra y tick style={grid=major, major grid style={dashed, semithick, black}},
  after end axis/.append code={
    \node[font=\tiny, anchor=north east, align=right, xshift=-0.5mm, yshift=-3mm]
      at (current axis.north east) {\tikz[baseline=-0.5ex]\draw[dashed, semithick] (0,0) -- (4mm,0);\ 10-ns limit\\(100\,MHz)};}]
\addplot[fill=dseBar, draw=none] coordinates {(12,42.3) (6,21.2) (4,13.3) (3,10.4) (2,8.7) (1,8.4)};
\addplot[fill=dseSel, draw=none] coordinates {(2,8.7)};

\nextgroupplot[title={(b) Latency (cycles)}, ymax=32, ytick={0,10,20,30}]
\addplot[fill=dseBar, draw=none] coordinates {(12,5) (6,7) (4,9) (3,11) (2,15) (1,27)};
\addplot[fill=dseSel, draw=none] coordinates {(2,15)};

\nextgroupplot[title={(c) LUTs ($\times 10^3$)}, ymax=9, ytick={0,3,6,9},
  xlabel={Rounds per cycle (MCP$r$)},
  every node near coord/.append style={/pgf/number format/.cd, fixed, precision=1}]
\addplot[fill=dseBar, draw=none] coordinates {(12,7.468) (6,4.340) (4,3.064) (3,2.290) (2,1.677) (1,1.061)};
\addplot[fill=dseSel, draw=none] coordinates {(2,1.677)};

\nextgroupplot[title={(d) Est.\ power, 5\,MHz (mW)}, ymin=120, ymax=270, ytick={120,170, 220,270},
  xlabel={Rounds per cycle (MCP$r$)}]
\addplot[fill=dseBar, draw=none] coordinates {(12,249) (6,201) (4,183) (3,174) (2,168) (1,166)};
\addplot[fill=dseSel, draw=none] coordinates {(2,168)};
\end{groupplot}
\end{tikzpicture}
\caption{Folded Ascon implementations executing $r$ rounds per cycle
(\texttt{MCP}$r$): (a)~critical path, w/10 ns limit (100~MHz);
(b)~cycles per authenticated operation, $24/r+3$; (c)~LUTs;
(d)~estimated power at 5~MHz. Only \texttt{MCP2} and \texttt{MCP1} meet
timing; we select \texttt{MCP2} (highlighted), the faster of the two.}
\Description{Four bar charts over rounds per cycle 12, 6, 4, 3, 2 and 1.
Critical path: 42.3, 21.2, 13.3, 10.4, 8.7 and 8.4 ns, with a dashed line
at 10 ns. Latency: 5, 7, 9, 11, 15 and 27 cycles. LUTs: 7468, 4340, 3064,
2290, 1677 and 1061. Estimated power at 5 MHz: 153, 105, 87, 78, 72 and
70 mW. MCP2 is highlighted.}
\label{fig:ascon-dse}
\end{figure}
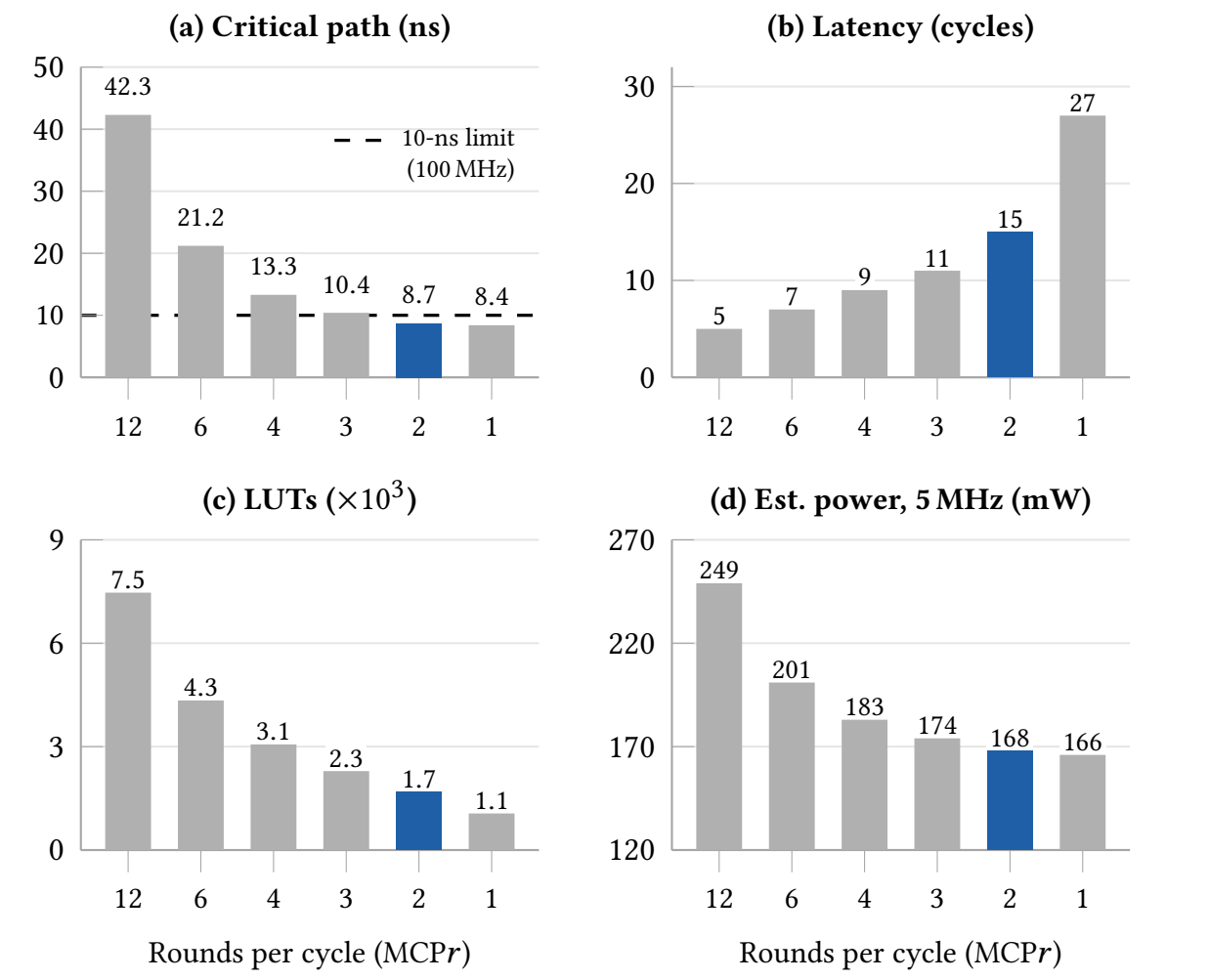

The folded variants reuse their round logic across cycles, executing $r=12,6,4,3,2,1$ rounds per cycle in designs denoted \texttt{MCP12} through \texttt{MCP1}. Their complete-operation latency is $24/r+3$ cycles: $24/r$ for the two permutations and three for control and data processing. Figure~\ref{fig:ascon-dse} reports these latencies alongside critical path, LUT utilisation, and estimated power. The module-level power comparison uses a common 5-MHz clock and identical estimation settings so that every variant can be evaluated, including those that cannot operate at 100~MHz. Only \texttt{MCP2} and \texttt{MCP1} meet the 10~ns period, with critical paths of 8.7 and 8.4~ns, respectively. Of these two, we select \texttt{MCP2} for its shorter operation: it takes 15 cycles, compared with 27 for \texttt{MCP1}, but uses 1,677 LUTs compared with 1,061 and has a 168-mW power estimate compared with 166~mW at 5~MHz. The selection favours latency; \texttt{MCP1} uses less logic and may suit a more restrictive area budget.

\subsection{Functional and Failure Tests}

RTL simulations were employed to test the acquisition transaction from the ADXL362 command and address sequence through byte reception, signed 12-bit reconstruction, sign extension, and X/Y packing. They also cover Ascon-128a encryption and decryption, atomic packet publication, and \texttt{VALID}/\texttt{ACK} handling, while separate ownership tests exercise safe and unsafe handover, irreversibility until reset, signal multiplexing, and MISO isolation. On the FPGA, the validation firmware first submits packets with one or more altered tag bits and observes a zero result with \texttt{DONE}=1 and \texttt{AUTH\_OK}=0. It then invokes decryption with an incomplete context, which returns zero immediately with \texttt{CONTEXT\_ERROR} and does not launch the decryptor. Processing a valid packet after each failed attempt checks that rejection leaves the CFU able to resume normal operation. Our validation covers these acceptance and rejection paths through directed tests, although the implementation has not undergone formal verification of the kind provided for VERSA's sensing monitor~\cite{ref08}.

\subsection{End-to-End Latency}

The ILA records registered one-cycle timing pulses, all sampled in the 100-MHz system-clock domain, so each stage latency can be obtained from the difference between its start and completion indices. Table~\ref{tab:latency} uses these markers to follow one accepted request through acquisition, encryption, software transport, and CFU completion. Timing from request acceptance excludes the work needed to issue the request and allows the measured hardware and software stages to be accounted for separately.

\begin{table}[t]
\caption{Latency from an accepted request to authenticated CFU completion at 100~MHz. Encryption occupies 15 of the 18 wrapper cycles; decryption occupies 15 of the 17 CFU cycles.}
\label{tab:latency}
\centering
\small
\setlength{\tabcolsep}{3pt}
\begin{tabularx}{\columnwidth}{@{}Xrr@{}}
\toprule
Stage & Cycles & Time ($\mu$s) \\
\midrule
SPI acquisition & 4,901 & 49.01 \\
Encryption, control, publication & 18 & 0.18 \\
Packet transport and CFU setup & 255 & 2.55 \\
CFU decryption and verification & 17 & 0.17 \\
\midrule
Total & 5,191 & 51.91 \\
\bottomrule
\end{tabularx}
\end{table}

Reading the four ADXL362 X/Y bytes takes 4,901 cycles ($49.01\,\mu\mathrm{s}$), after which the \texttt{MCP2} engine encrypts the sample in 15 cycles ($0.15\,\mu\mathrm{s}$). Wrapper control and packet publication bring the interval from accepted \texttt{START} to \texttt{packet\_valid} to 4,919 cycles ($49.19\,\mu\mathrm{s}$). Software then copies the sequence, ciphertext, and tag, acknowledges the packet, loads the CFU context, and issues \texttt{AUDEC}. These operations occupy 255 cycles ($2.55\,\mu\mathrm{s}$) between publication and instruction acceptance. The CFU requires a further 17 cycles ($0.17\,\mu\mathrm{s}$), including 15 in the decryption engine, to complete authentication. Summing the stages gives 5,191 cycles ($51.91\,\mu\mathrm{s}$) from accepted acquisition to authenticated CFU completion, when \texttt{AUTH\_OK} is asserted for a genuine packet. Sensor conversion and filtering, as well as software before acceptance or after completion, are outside this measured interval.

Acquisition accounts for 4,901 of the 5,191 cycles (94.4\%), compared with 30 cycles across the two cryptographic engines. Even the 255-cycle transfer and context-loading interval is much longer than the cryptographic processing itself. Reducing permutation latency further would consequently have little effect on the complete path unless acquisition or software transport also changed. Although \texttt{MCP2} is the faster of the two engines that meet the clock constraint, the area savings of \texttt{MCP1} may be preferable in a system where this small contribution to total latency can be increased.

\subsection{Sustained Throughput}

To measure transaction capacity, the firmware executes 10,000 acquisitions consecutively, with no artificial delays or UART output during the timed interval. Each transaction includes sensor acquisition, Ascon-128a encryption, packet transfer and acknowledgement, CFU-context loading, and authenticated decryption. The run completes in 52,590,015 cycles at 100~MHz, giving an average interval of 5,259 cycles ($52.59\,\mu\mathrm{s}$) and a rate of 19,015 transactions/s. No acquisition failures, authentication failures, sequence discontinuities, or overruns occur. An ILA spot check records three consecutive \texttt{start\_accepted\_dbg} pulses at sample indices 256, 5,515, and 10,774. Both gaps are 5,259 cycles, agreeing with the firmware counter and showing the same spacing for the captured transactions. The sensor's 100-Hz output-data rate limits how often those transactions can obtain a new physical measurement, so 19,015 transactions/s is the capacity of the protected path rather than a sample rate. At 100~Hz, a protected acquisition occupies 5,259 of every 1,000,000 processor cycles (0.53\%), of which the two cryptographic engines account for 30. 

\subsection{Comparison with Software SPI}

The baseline acquires the same four X/Y bytes through NEORV32's standard SPI peripheral using blocking software transfers, then reconstructs, sign-extends, and packs them for the application. It uses the same 100-MHz processor, approximately 1.04-MHz SPI clock, and sensor configuration. Its response interval is measured from immediately before software requests chip-select assertion until the packed sample is available to the application. Across 10,000 measurements, that interval ranges from 5,512 to 5,568 cycles and averages 5,519 cycles ($55.19\,\mu\mathrm{s}$). A separate continuous run takes 55,379,991 cycles for 10,000 transactions, corresponding to a mean interval of approximately 5,538 cycles ($55.38\,\mu\mathrm{s}$), 18,057 transactions/s, and 0.55\% of processor cycles at 100~Hz (Table~\ref{tab:comparison}).

\begin{figure}[t]
\centering
\definecolor{ltBase}{RGB}{176,176,176}
\definecolor{ltAcq}{RGB}{31,95,168}
\definecolor{ltEnc}{RGB}{230,145,56}
\definecolor{ltPub}{RGB}{140,90,170}
\definecolor{ltSw}{RGB}{120,170,220}
\definecolor{ltCfu}{RGB}{60,150,90}
\definecolor{ltSG}{RGB}{200,40,40}
\begin{tikzpicture}[
  call/.style={font=\scriptsize, inner sep=0.8pt, anchor=south},
  lead/.style={black!60, very thin},
]
\begin{groupplot}[
  group style={group size=1 by 3, vertical sep=11mm},
  scale only axis, width=56mm,
  xbar stacked, /pgf/bar width=4mm,
  ytick={0,1}, yticklabels={Baseline, \schemename{}},
  xmin=0, xmax=6000, xtick={0,2000,...,6000}, minor xtick={1000,3000,5000},
  minor grid style={gray!20}, xminorgrids,
  scaled x ticks=false,
  xticklabel style={/pgf/number format/.cd, 1000 sep={\mbox{,}}},
  tick label style={font=\footnotesize},
  title style={font=\footnotesize\bfseries, yshift=-0.8ex},
  axis line style={gray!70}, tick style={gray!70}, ytick style={draw=none},
  xmajorgrids, grid style={gray!20},
  axis y line*=left, axis x line*=bottom,
  every axis plot/.append style={draw=white, line width=0.3pt},
  clip=false,
]

\nextgroupplot[title={(a) Response latency (cycles)},
  height=27mm, ymin=-0.6, ymax=2.15,
  legend to name=ltlegend, legend columns=3,
  legend style={font=\footnotesize, draw=none,
    /tikz/every even column/.append style={column sep=1.2mm}},
  legend cell align=left, area legend]
\addplot[fill=ltBase] coordinates {(5519,0) (0,1)};
\addlegendimage{area legend, fill=ltSG, draw=white}
\addplot[fill=ltAcq]  coordinates {(0,0) (4901,1)};
\addplot[fill=ltEnc]  coordinates {(0,0) (15,1)};
\addplot[fill=ltPub]  coordinates {(0,0) (3,1)};
\addplot[fill=ltSw]   coordinates {(0,0) (255,1)};
\addplot[fill=ltCfu]  coordinates {(0,0) (17,1)};
\legend{Baseline, \schemename{}, Acquisition, Encryption,
        Publication, SW transfer, CFU verify}

\coordinate (acqMid) at (axis cs:2450,1);
\coordinate (convMid) at (axis cs:2760,0);
\coordinate (sEnd) at (axis cs:5191,1);
\coordinate (cEnd) at (axis cs:5519,0);
\coordinate (encT) at (axis cs:4908,1.27);
\coordinate (pubT) at (axis cs:4917.5,1.27);
\coordinate (swT)  at (axis cs:5046,1.27);
\coordinate (cfuT) at (axis cs:5182.5,1.27);
\coordinate (encL) at (axis cs:3500,1.45);
\coordinate (cfuL) at (axis cs:5950,1.45);
\coordinate (pubL) at (axis cs:3800,1.80);
\coordinate (swL)  at (axis cs:5550,1.80);
\coordinate (dA) at (axis cs:5191,0.5);
\coordinate (dB) at (axis cs:5519,0.5);

\nextgroupplot[title={(b) Steady-state interval (cycles)},
  height=14mm, ymin=-0.6, ymax=1.6]
\addplot[fill=ltBase] coordinates {(5538,0) (0,1)};
\addplot[fill=ltSG]   coordinates {(0,0) (5259,1)};
\coordinate (bS) at (axis cs:5259,1);
\coordinate (bC) at (axis cs:5538,0);

\nextgroupplot[title={(c) Sustained throughput (transactions/s)},
  height=14mm, ymin=-0.6, ymax=1.6,
  xmin=0, xmax=20000, xtick={0,10000,20000},
  minor xtick={5000,15000}]
\addplot[fill=ltBase] coordinates {(18057,0) (0,1)};
\addplot[fill=ltSG]   coordinates {(0,0) (19015,1)};
\coordinate (tS) at (axis cs:19015,1);
\coordinate (tC) at (axis cs:18057,0);
\end{groupplot}

\node[font=\scriptsize, text=white] at (acqMid) {Acquisition 4,901};
\node[font=\scriptsize] at (convMid) {CPU-driven SPI + software};
\node[font=\scriptsize, anchor=west] at (sEnd) {5,191};
\node[font=\scriptsize, anchor=west] at (cEnd) {5,519};
\node[call] (cEnc) at (encL) {Encrypt 15};
\draw[lead] (cEnc.south) -- (encT);
\node[call] (cCfu) at (cfuL) {CFU 17};
\draw[lead] (cCfu.south) -- (cfuT);
\node[call] (cPub) at (pubL) {Publish 3};
\draw[lead] (cPub.south) -- (pubT);
\node[call] (cSw) at (swL) {SW 255};
\draw[lead] (cSw.south) -- (swT);
\draw[|-|, thin] (dA) -- (dB);
\node[font=\scriptsize, anchor=east] at ([xshift=-1mm]dA) {$-$328 cycles ($-$5.9\%)};
\node[font=\scriptsize, anchor=east, text=white, xshift=-1mm] at (bS) {5,259 ($-$5.0\%)};
\node[font=\scriptsize, anchor=east, xshift=-1mm] at (bC) {5,538};
\node[font=\scriptsize, anchor=east, text=white, xshift=-1mm] at (tS) {19,015 (+5.3\%)};
\node[font=\scriptsize, anchor=east, xshift=-1mm] at (tC) {18,057};
\node[anchor=south] at ([yshift=5.5mm, xshift=-10mm]group c1r1.north)
  {\pgfplotslegendfromname{ltlegend}};
\end{tikzpicture}
\caption{Latency of \schemename{} sensing transactions
(vs.~blocking-SPI baseline at 100\,MHz). (a)~Response latency from acceptance of the
acquisition request to authenticated plaintext release. (b)~Average
steady-state interval over 10,000 transactions. (c)~Throughput.}
\Description{Three horizontal bar charts.
(a) Response latency: \schemename{} 5,191 cycles, made of acquisition 4,901,
encryption 15, wrapper and publication 3, software-mediated transfer 255,
and CFU decryption and verification 17; Baseline 5,519 cycles.
\schemename{} is 328 cycles (5.9 percent) faster. (b) Steady-state interval:
\schemename{} 5,259 cycles; Baseline 5,538 cycles. (c) Throughput:
\schemename{} 19,015 transactions per second; Baseline 18,057 transactions per
second, a 5.3 percent increase.}
\label{fig:latency}
\end{figure}
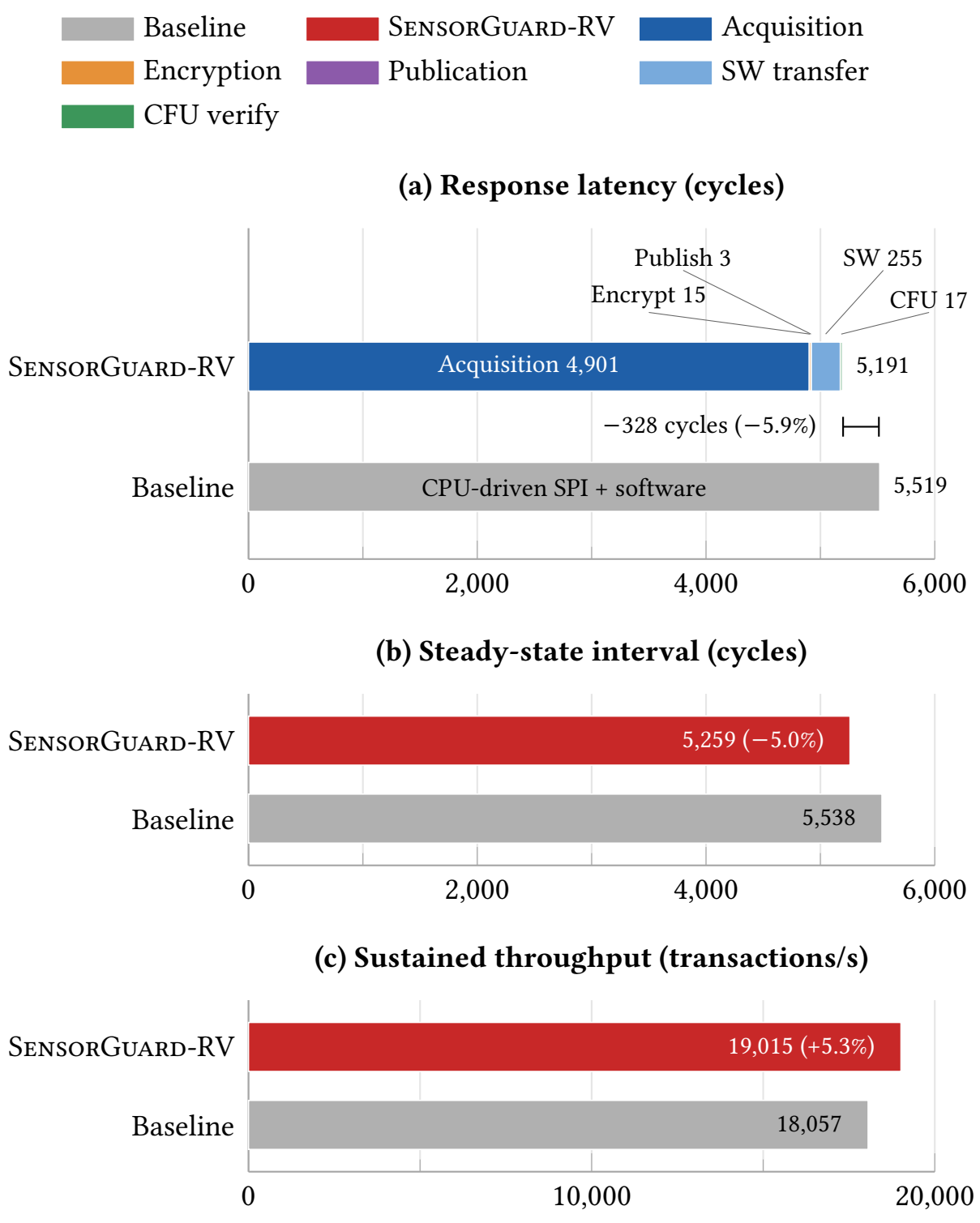

The baseline's 5,519-cycle response interval exceeds the 5,191-cycle interval of \schemename{} by 328 cycles ($3.28\,\mu\mathrm{s}$), or 5.94\%, but the timing boundaries differ. The protected path starts at hardware request acceptance and ends at CFU completion, whereas the baseline includes software before chip-select assertion and ends with an application-visible sample. The difference cannot therefore be taken as a matched application-level latency reduction. In the continuous runs, \schemename{} reduces the mean transaction interval by approximately 279 cycles ($2.79\,\mu\mathrm{s}$), increasing throughput by 5.31\% while including packet handling and both 15-cycle cryptographic operations. Dedicated acquisition removes the per-byte software polling and memory-mapped-I/O sequence used by the baseline, which can account for a faster complete loop despite the added protection. The experiment changes acquisition and cryptography together, however; measuring an otherwise identical fixed-function controller without cryptography would be necessary to determine their individual contributions. Figure ~\ref{fig:latency} compares the latency of one protected transaction and the average steady-state interval over 10,000 transactions in \schemename{} versus the blocking-SPI baseline, at 100 MHz.

\begin{table}[t]
\caption{Software-driven baseline and \schemename{} at 100~MHz. Throughput is measured over 10,000 consecutive transactions; resources and power come from non-instrumented post-implementation builds, and both designs meet the 10-ns timing constraint.}
\label{tab:comparison}
\centering
\small
\setlength{\tabcolsep}{3pt}
\begin{tabularx}{\columnwidth}{@{}Xrrr@{}}
\toprule
Metric & Baseline & \schemename{} & Change \\
\midrule
Mean loop interval (cycles) & 5,538 & 5,259 & $-$5.0\% \\
Throughput (transactions/s) & 18,057 & 19,015 & $+$5.3\% \\
\midrule
LUTs & 1,954 & 5,104 & $+$161.2\% \\
Registers & 1,517 & 3,751 & $+$147.3\% \\
Est.\ power (mW) & 118 & 368 & $+$211.9\% \\
\bottomrule
\end{tabularx}
\end{table}

\subsection{Implementation Cost}

Table~\ref{tab:comparison} also compares the hardware cost of the complete protected system, with \texttt{MCP2} encryption and decryption engines, against the conventional NEORV32 baseline. Both are post-implementation builds without ILA instrumentation and use the same memory capacities, target device, clock constraint, and Vivado estimation settings. Processor parameters are unchanged except where the \schemename{} extensions require additional hardware, so the reported differences cover the acquisition wrapper, ownership control, packet interface, and CFU together.

Total LUT utilisation rises from 1,954 to 5,104 (+161.2\%) and slice-register utilisation from 1,517 to 3,751 (+147.3\%). The SensorGuard-RV modules themselves account for 3,155 LUTs and 2,215 registers; the totals differ slightly from these sums because re-implementation changes the common SoC logic (1,954 to 1,949 LUTs and 1,517 to 1,536 registers). These modules occupy 5.0\% of the XC7A100T’s 63,400 LUTs and 1.7\% of its 126,800 flip-flops. Vivado-estimated full-system power rises from 118 to 368~mW, an increase of 250~mW (211.9\%). Both designs meet the 10-ns timing constraint, so these costs are incurred without reducing the baseline 100-MHz operating frequency. We did not implement a TEE or tagged-bus alternative on this platform, so the measured increases support a cost comparison with conventional acquisition only.

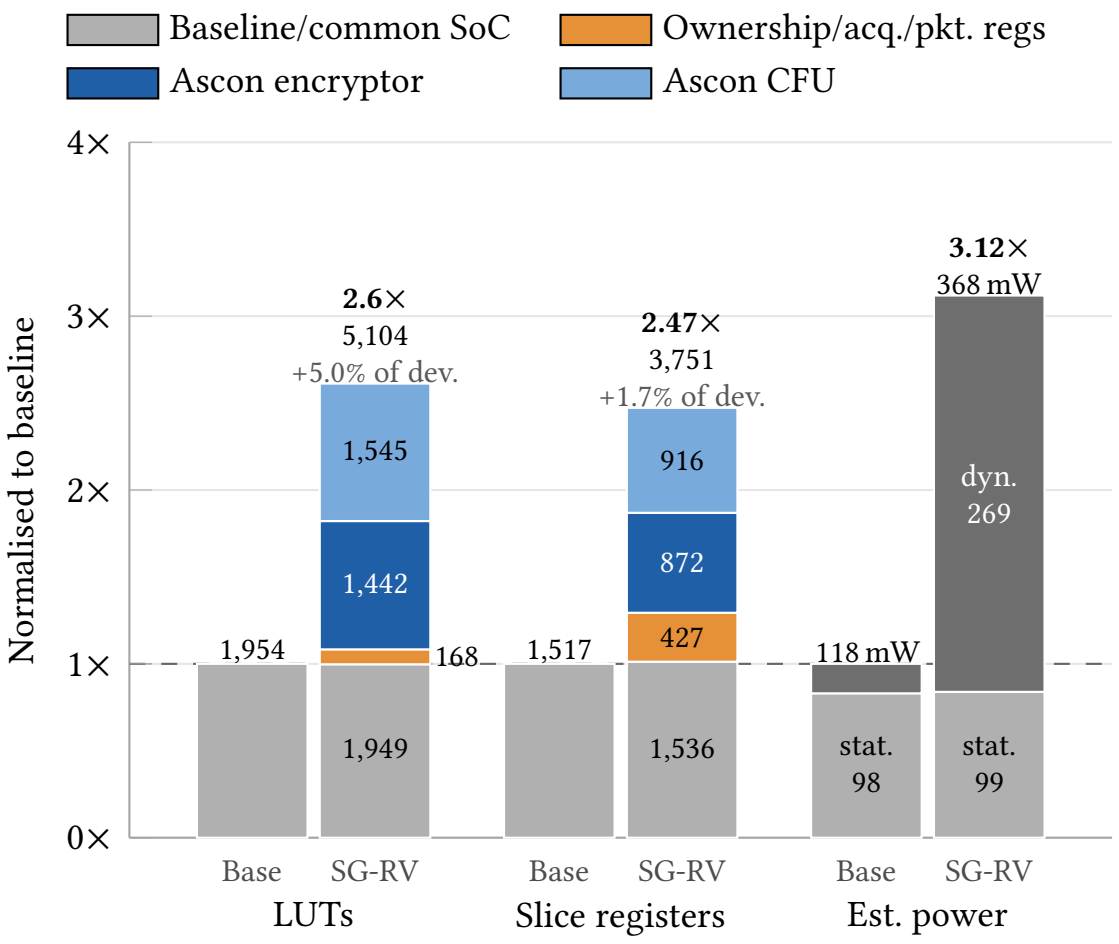
\begin{figure}[t]
\centering
\definecolor{nmBase}{RGB}{176,176,176}  
\definecolor{nmDyn}{RGB}{110,110,110}   
\definecolor{nmCtrl}{RGB}{230,145,56}   
\definecolor{nmEnc}{RGB}{31,95,168}     
\definecolor{nmDec}{RGB}{120,170,220}   
\begin{tikzpicture}[
  seg/.style={draw=white, line width=0.4pt},
  segl/.style={font=\scriptsize, inner sep=0pt},
]
\begin{axis}[
  scale only axis, width=65mm, height=46mm,
  xmin=0.4, xmax=3.6,
  xtick={1,2,3},
  xticklabels={LUTs, Slice registers, Est.\ power},
  xticklabel style={yshift=-2.3ex},
  ymin=0, ymax=4, ytick={0,1,2,3,4},
  ylabel={Normalised to baseline},
  ylabel style={font=\footnotesize},
  tick label style={font=\footnotesize},
  yticklabel={\pgfmathprintnumber{\tick}$\times$},
  axis line style={gray!70}, tick style={gray!70},
  ymajorgrids, grid style={gray!20},
  axis x line*=bottom, axis y line*=left,
  x tick style={draw=none},
  extra y ticks={1}, extra y tick labels={},
  extra y tick style={grid=major, major grid style={dashed, black!60}},
  clip=false,
  legend style={font=\footnotesize, draw=none, at={(0.44,1.03)}, anchor=south,
                legend columns=2,
                /tikz/every even column/.append style={column sep=2.5mm}},
  legend cell align=left,
]
\addlegendimage{area legend, fill=nmBase, draw=none}
\addlegendentry{Baseline/common SoC}
\addlegendimage{area legend, fill=nmCtrl, draw=none}
\addlegendentry{Ownership/acq./pkt.\ regs}
\addlegendimage{area legend, fill=nmEnc, draw=none}
\addlegendentry{Ascon encryptor}
\addlegendimage{area legend, fill=nmDec, draw=none}
\addlegendentry{Ascon CFU}

\fill[nmBase, seg] (axis cs:0.62,0) rectangle (axis cs:0.98,1);
\fill[nmBase, seg] (axis cs:1.02,0)           rectangle (axis cs:1.38,1949/1954);
\fill[nmCtrl, seg] (axis cs:1.02,1949/1954)   rectangle (axis cs:1.38,2117/1954);
\fill[nmEnc,  seg] (axis cs:1.02,2117/1954)   rectangle (axis cs:1.38,3559/1954);
\fill[nmDec,  seg] (axis cs:1.02,3559/1954)   rectangle (axis cs:1.38,5104/1954);
\node[segl, anchor=south] at (axis cs:0.80,1) {1,954};
\node[segl, anchor=south, align=center] at (axis cs:1.20,5104/1954)
  {\textbf{2.6$\times$}\\5,104\\{\color{black!65}+5.0\% of dev.}};
\node[segl] at (axis cs:1.20,0.50) {1,949};
\node[segl, anchor=west] at (axis cs:1.39,2033/1954) {168};
\node[segl, text=white] at (axis cs:1.20,2838/1954) {1,442};
\node[segl] at (axis cs:1.20,4331.5/1954) {1,545};

\fill[nmBase, seg] (axis cs:1.62,0) rectangle (axis cs:1.98,1);
\fill[nmBase, seg] (axis cs:2.02,0)           rectangle (axis cs:2.38,1536/1517);
\fill[nmCtrl, seg] (axis cs:2.02,1536/1517)   rectangle (axis cs:2.38,1963/1517);
\fill[nmEnc,  seg] (axis cs:2.02,1963/1517)   rectangle (axis cs:2.38,2835/1517);
\fill[nmDec,  seg] (axis cs:2.02,2835/1517)   rectangle (axis cs:2.38,3751/1517);
\node[segl, anchor=south] at (axis cs:1.80,1) {1,517};
\node[segl, anchor=south, align=center] at (axis cs:2.20,3751/1517)
  {\textbf{2.47$\times$}\\3,751\\{\color{black!65}+1.7\% of dev.}};
\node[segl] at (axis cs:2.20,0.50) {1,536};
\node[segl] at (axis cs:2.20,1749.5/1517) {427};
\node[segl, text=white] at (axis cs:2.20,2399/1517) {872};
\node[segl] at (axis cs:2.20,3293/1517) {916};

\fill[nmBase, seg] (axis cs:2.62,0)      rectangle (axis cs:2.98,98/118);
\fill[nmDyn,  seg] (axis cs:2.62,98/118) rectangle (axis cs:2.98,1);
\fill[nmBase, seg] (axis cs:3.02,0)      rectangle (axis cs:3.38,99/118);
\fill[nmDyn,  seg] (axis cs:3.02,99/118) rectangle (axis cs:3.38,368/118);
\node[segl, anchor=south east] at (axis cs:2.98,1) {118\,mW};
\node[segl, anchor=south, align=center] at (axis cs:3.20,368/118)
  {\textbf{3.12$\times$}\\368\,mW};
\node[segl, align=center] at (axis cs:2.80,0.42) {stat.\\98};
\node[segl, align=center] at (axis cs:3.20,0.42) {stat.\\99};
\node[segl, align=center, text=white] at (axis cs:3.20,1.98) {dyn.\\269};

\pgfplotsinvokeforeach{0.80,1.80,2.80}{
  \node[font=\scriptsize, anchor=north, text=black!70, yshift=-1pt] at (axis cs:#1,0) {Base};}
\pgfplotsinvokeforeach{1.20,2.20,3.20}{
  \node[font=\scriptsize, anchor=north, text=black!70, yshift=-1pt] at (axis cs:#1,0) {SG-RV};}
\end{axis}
\end{tikzpicture}
\caption{\schemename{} (SG-RV) post-implementation cost normalised to
the NEORV32 baseline (dashed line).
}
\Description{Grouped bar chart comparing SenseGuard-RV with the
conventional baseline, normalised to the baseline. LUTs: 5,104 versus
1,954 (2.61 times), the SenseGuard-RV bar stacked as 1,949 common SoC,
168 ownership, acquisition and packet registers, 1,442 Ascon encryptor
and 1,545 Ascon CFU decryptor. Slice registers: 3,751 versus 1,517 (2.47
times), stacked as 1,536 common, 427 control, 872 encryptor and 916 CFU.
Estimated power: 368 mW versus 118 mW (3.12 times), shown as stacked
bars: baseline 98 mW static plus 20 mW dynamic; SenseGuard-RV 99 mW
static plus 269 mW dynamic.}
\label{fig:impl-norm}
\end{figure}

Figure ~\ref{fig:impl-norm} compares \schemename{}'s post-implementation cost with the conventional NEORV32 baseline by component. The complete system uses 5,104
LUTs and 3,751 slice registers, $2.60\times$ and $2.47\times$ the
baseline's 1,954 LUTs and 1,517 registers. Of the 3,155
LUTs and 2,215 registers in these additional modules, the two Ascon-128a engines account for 2,987 LUTs
(94.7\%): 1,442 LUTs (45.7\%) for the sensor-side encryptor and 1,545 LUTs (49.0\%) for the CFU, and 1,788 registers (80.7\%):  872 registers (39.4\%) for the sensor-side encryptor and 916 registers (41.4\%) for the CFU, which
also holds the tag context and the verification and release logic. The
security mechanisms specific to \schemename{}, namely the ownership
switch, the fixed-function acquisition controller, and the packet
registers, add only 168 LUTs (5.3\% of added LUTs) and 427 registers (19.3\% of added registers), most of the latter being
the 192-bit packet buffer and its status. The cost of \schemename{}
therefore lies mainly in its cryptographic primitives, not in its
ownership and acquisition control, and any reduction in Ascon
implementation cost would carry over directly. The complete system uses 8.1\% and 3.0\% of the device's
LUTs and registers, respectively. Device static power is practically identical (98
vs.\ 99\,mW), so the increase is entirely dynamic (20 vs.\ 269\,mW).

\subsection{Limitations}
\label{sec:limitations}

Our prototype disables NEORV32's DMA controller, requiring software to copy each protected packet from the wrapper to data memory and then load it into the CFU. A controlled DMA engine could perform the first transfer without exposing the packet's plaintext, but CFU loading and invocation would still require processor instructions because its inputs arrive through register operands. A DMA master could corrupt, delete, or delay them; with access to wrapper MMIO, it can also acknowledge packets prematurely or trigger unwanted acquisitions. Such a master could also observe plaintext that the consumer stores after verification. Consequently, deployments should strongly consider an I/O physical memory protection (IOPMP) unit in order to restrict DMA access.

Applications that need to reconfigure a sensor or protect computations on its output would require changes to the present design. Fortress retains trusted drivers and application transformations~\cite{ref07}, and VERSA authorises the sensing computation~\cite{ref08}; both support forms of protected software execution that our fixed acquisition path does not provide.
In our prototype, the key is fixed in the FPGA configuration. A realistic deployment would protect it with bitstream encryption or derive the key (e.g., from a PUF~\cite{ref09}), and would need per-device key provisioning. 

Lastly, supporting multiple sensors would require per-source acquisition front ends scheduled by trusted hardware, bounded queues with overflow reporting, and source, format, and timestamp fields authenticated as associated data~\cite{ascon-v12}. Nonces would need disjoint per-source counters under the per-boot epoch, the CFU would need a sequence register for each source, and larger payloads would require multi-block support in both engines and the CFU.  Furthermore, several consumers would need to serialise access to the shared CFU context; sharing buses and engines would add queueing delay to be measured against application deadlines. Board-level power measurements and a side-channel evaluation would further evidence implementation cost and attack resistance outside the present model. We defer these avenues to future work.

\section{Conclusion}
\label{sec:conclusion}
This work presented \schemename{}, showing that a RISC-V microcontroller-class system can deliver a sensor sample from the SPI bus to a processor register
without trusting intermediate software, buses, or memory. Our central idea is to move the decision to release a sample into hardware: once a fixed-function
controller owns the sensor and the processor writes back only samples whose tags verify, components between the two can remain untrusted. The cost of this falls principally on the cipher: the Ascon engines account for 30 of the 5,191 cycles needed to bring an accelerometer sample into a register, and the protected loop still runs faster than software-driven SPI. The ownership switch, acquisition controller, and packet registers together add only 168 LUTs; the rest of the added logic is the two Ascon engines, so improvements in compact Ascon hardware would carry over directly. Two extensions would move the design towards deployment, which are left as future work. Firstly, bounding the time at which an old sampled can be released would enable a consumer reject delayed readings as well as replayed ones. Secondly, supporting several sensors and larger payloads requires per-source sequence state and authenticated
metadata.

\begin{acks}
This work was supported by the UK Engineering and Physical Sciences Research Council via the ‘Chameleon’ project (EP/Y030168/1).
\end{acks}

\printbibliography

\end{document}